# Projection Of Temperature And Precipitation For 2020-2100 For Tehran Region Using Post-processing Of General Circulation Models Output And Artificial Neural Network Approach

Ehsan Mosadegh[1,*], Iman Babaeian[2]

[1] Atmospheric Sciences Department, University of Nevada, Reno, NV 89557, USA.

[2] Climate Change Division, Climate Research Institute, Mashhad, Iran.

## Abstract

Multi-model projections in climate studies are performed to quantify uncertainty and improve reliability in climate projections. The challenging issue is that there is no unique way to obtain performance metrics, nor is there any consensus about which method would be the best method of combining models. The goal of this study was to investigate whether combining climate model projections by artificial neural network (ANN) approach could improve climate projections and therefore reduce the range of uncertainty. The equally-weighted model averaging (the mean model) and single climate model projections (the best model) were also considered as references for the ANN combination approach. Simulations of present-day climate and future projections from 15 General Circulation Models (GCMs) for temperature and precipitation were employed. Results indicated that combining GCM projections by the ANN combination approach significantly improved the simulations of present-day temperature and precipitation than the best model and the mean model. The identity of the best model changed between the two variables and among stations. Therefore, there was not a unique model which could represent the best model for all variables and/or stations over the study region. The mean model was also not skillful in giving a reliable projection of historical climate. Simulation of temperature indicated that the ANN approach had the best skill at simulating present-day monthly means than other approaches in all stations. Simulation of present-day precipitation, however, indicated that the ANN approach was not the best approach in all stations although it performed better than the mean model. Multi-model projections of future climate conditions performed by the ANN approach projected an increase in temperature and reduction in precipitation in all stations and for all scenarios.

# 1. Introduction

General Circulation Models (GCMs) are considered important tools for simulating future global climate. These models are able to simulate different components of the Earth system such as the atmosphere and oceans. However, due to their coarse resolution, projections of these models have low confidence and high uncertainty. Furthermore, Using the output of a single GCM in climate change projections does not yield realistic projections of future climate conditions. Intercomparison studies of GCMs indicate that each climate model has different skills in simulating certain aspects of the climate system mechanisms (Lambert and Boer 2001; Gleckler et al, 2008). This means that climate variables are simulated with different degrees of accuracy by different models, and no single model delivers the best simulation for all variables and/or all regions. Therefore, in order to quantify the range of uncertainty in climate change projections, Intergovernmental Panel on Climate Change (IPCC) recommends using multiple GCMs in climate simulations (Parry et al. 2007).

Uncertainty in climate projections usually arises from three main sources: internal variability of the climate system, which stems from natural fluctuations of the climate without considering the effect of radiative forcing of the planet; emission scenarios, that stems from uncertainties in estimating future emissions of aerosols and greenhouse gases; and model errors, that stems from model formulations and structural uncertainties (Little et al. 2015). The domination of the three sources of uncertainty in climate change projections varies with spatial and temporal scales (Räisänen 2001; Cox and Stephenson 2007). For projections in the range of a decade or two, the dominant sources of uncertainty are model uncertainty and internal variability. In projections of longer time scales, such as climate change conditions until the end of the $21^{st}$ century, model uncertainty and scenario uncertainty become the dominant sources. Hawkins and Sutton (2009) showed that the importance of internal variability would increase at smaller spatial scales, and it was generally the dominant source in short-time scales projections. Their study also indicated that the importance of internal variability would decline when the projection time increased. Moreover, scenario uncertainty made an important contribution over many regions of the world at the end of the $21^{st}$ century. Based on their study, model uncertainty had an important role in both global and regional scales and made a significant contribution to all time scales.

Model uncertainty plays an important role in studying the climate change conditions for the next century. In order to address model uncertainty in climate change simulations, a multi-model combination has been adopted as a well-accepted approach which generally increases the reliability of model forecasts (Weigel et al. 2010). However, so far no consensus has been reached about which method would be the best method of combining outputs of several climate models. Generally, in climate change multi-model studies, a common practice is to use the concept of weighting the outputs of climate models (Tebaldi and Knutti 2007). The approaches fall into two general categories of *equal weighting* and *skill-based weighting*. Equal weighting is the easiest approach, in which every model is given equal weight regardless of its magnitude of contribution to the combination. *Skill-based* weighting is a more sophisticated approach. In this approach, every individual model is given a different weight based on their contribution to projections. The weights are calculated based on the skill of every individual model in simulating the present-day climate conditions and therefore are considered as skill-based weights. A study by (Giorgi and Mearns 2002) proposed the "Reliability Ensemble Averaging" (REA) approach to weight different models based on their contributions to present-day climate simulations. They defined two reliability criteria in multi-model studies to evaluate skills of GCMs in simulating climate variables in the present and future climates: "model performance" that indicated how well the models can simulate the baseline (present-day) climate, and "model convergence" that investigated the convergence between the simulations of future climate across models. The underlying philosophy of the REA approach in a multi-model projection is to detect models with weak performance in simulating present-day climate (the outliers) and to reduce their role in projections by assigning them less weight than models with small bias and good performance. However, it has been argued that because common weaknesses in the representation of certain climate processes may exist among a majority of models, outliers may not appear at random. Therefore, considering and analyzing a subset of models as the best guess, whose agreement is considered as their individual tendencies, may result in disregarding the possible range of uncertainty in the convergence criterion.

Multi-model combination based on model weighting has been the focus of multiple studies (Lambert and Boer 2001; Tebaldi et al. 2005; Min and Hense 2006; Gleckler et al., 2008; Weigel et al. 2010; Hesselbjerg Christensen et al. 2010; Reto Knutti et al. 2009; Arzhanov et al., 2012). A study by Lambert and Boer (2001) indicated that no one model is best for all variables and/or all regions, and different variables are simulated with different levels of success by different models. They also concluded that the equally-weighted average or the "mean model" usually provides the best comparison to observations than the single models. A similar study was conducted by Gleckler et al., (2008) which emphasized the results of Lambert and Boer (2001). They ranked models based on simulating each variable that was considered in their study and concluded that the ranking of models varied from one variable to the other one. They also considered the mean model in their study and demonstrated that the mean model would outperform all single models in nearly every aspect. In a multi-model study based on model weighting, Weigel et al. (2010) suggested that equally weighted multi-model on average would outperform single-model projections. They also considered the weighting of models and demonstrated that if the optimum weighting of the models were accurately performed, projection errors would be reduced in simulations. On the other hand, if inappropriate weights were assigned, which did not represent the skill of the model, the weighted multi-model would perform on average worse than equally weighted models, and therefore more information would be lost than were supposed to acquire from simulations. The task of assigning weights to models is performed by defining some metrics to quantify model performance. The difficulty in this procedure is that there is no unique way to obtain metrics, nor is there any consensus about which method would be the best method of combining models. This difficulty is highlighted by the fact that the choice of the metrics to weight models is a pragmatic and subjective task that may incorporate more uncertainty into projections (Tebaldi and Knutti 2007; Hesselbjerg Christensen et al. 2010).

In this study, we investigate an alternative modeling approach to combine multiple climate change projections. We combined outputs of several GCMs with an artificial neural network (ANN) model to obtain a multi-model combination. The purpose of the suggested approach is to investigate how much the combination of GCM projections by using our ANN approach would improve multi-model projections and therefore could reduce the uncertainty by obtaining an optimal models combination. In order to assess the results, projections from two common approaches namely single climate models (the best model) and equal weighting of the models (the mean model) were compared with this approach. The ANN approach derives an optimal combination of multiple climate models by correlating the GCM simulations at grid-scale to observations of climate variables at the local scale. This procedure can benefit the climate projections because it reduces the subjectivity and complexity aspects in constructing and combining metrics used for weighting the models.

Climate change is projected to impact each component of the climate system with regional differences (IPCC, 2021; Mejia et al., 2019; Mosadegh et al., 2018; Mosadegh and Nolin, 2020). A few studies have addressed the uncertainty of climate projections over the 21st century for the Tehran region (Mosadegh et al., 2013; Mosadegh and Babaeian, 2021) and have investigated to what extent the projected changes in climate variables can affect other aspects of our environment such as air pollution (Mosadegh et al., 2013; Mosadegh et al., 2021). In order to investigate the skill of the suggested ANN approach, we simulated temperature and precipitation for the study region. Moreover, we used the ANN approach to obtain a multi-model projection of temperature and precipitation for the future climate change conditions of the study region to the end of the $21^{st}$ century. In this projection, the focus was mainly on the projection aspect itself rather than the model convergence criterion, and also to know to what extent this approach can reduce the uncertainty in projections. We also took two sources of uncertainty into consideration: model uncertainty and scenario uncertainty. Internal climate variability is often considered negligible on long time scales (Hawkins and Sutton 2009).

This paper is structured as follows: Data, models, and scenarios used in the present study are described in section 2. The employed methodology is presented in section 3. The results are discussed in section 4, and conclusions are provided in section 5.

## 2. Data, models, and scenarios

### 2.1. Data

Observation data sets from four synoptic stations in Tehran province in the Alborz mountain domain were used in this study. In each station, available long-term observations of monthly surface temperature and precipitation until the year 2000 were used as the baseline period. For training the ANN, long-term observation data sets were necessary. Therefore, the baseline period for each station was selected based on the availability of observed data until the year 2000. Monthly surface temperature data sets for every station were obtained from daily observed values in each station. The precipitation data sets were obtained from daily observed values in each station and then were totaled to obtain the total monthly precipitation in each station. The information of stations is given in table 1. In the present study, calculations for handling large data sets and obtaining the indices were coded in MATLAB.

Table 1 Information of station used in the present study

| Station | Latitude | Longitude | Elevation (m) | Baseline |
|---|---|---|---|---|
| Karaj | 35 55 N | 50 54 E | 1312.5 | 1985-2000 |
| Mehrabad | 35 41 N | 51 19 E | 1190.8 | 1960-2000 |
| Doshan Tappeh | 35 42 N | 51 20 E | 1209.2 | 1972-2000 |
| Abali | 35 45 N | 51 53 E | 2465.2 | 1983-2000 |

### 2.2. Models

Uncertainty from Climate models is an important source of uncertainty in both short-term and long-term climate change projections (Räisänen 2001; Hawkins and Sutton 2009). Every model is skillful in capturing some aspects of the climate system and there is no one model that is skillful to simulate all variables and/or regions (Lambert and Boer 2001). We focused our work on a multi-model combination analysis and used the maximum available number of climate models in order to consider the widest possible range of model projections. This enables us to consider the skills of all models together in projections. The GCM projections for both historical and future climate conditions were obtained from Canadian Climate Data and Scenarios database (http://ccds-dscc.ec.gc.ca) for the study region. This database provides monthly simulations of a broad range of climate variables based on different emission scenarios (A1B, A2, B1) and geographical position of every location from 24 Atmospheric-Oceanic General Circulation Models. The employed simulations were from a subset of GCMs which were used in the IPCC 4$^{th}$ assessment report/CMIP3. We only considered GCMs that would provide projections of all three emission scenarios for the study region in order to consider the uncertainty from emissions scenarios in projections. In order to use the maximum number of GCMs in our multi-model projections, we selected 15 models which provided all three simulations of A1B, A2, and B1 emission scenarios from the set of 24 GCMs. The list of employed GCMs is given in table 2.

Table 2 Features of the GCMs from IPCC AR4 used in this study

| Country | Developer | GCM | Model acronym | Grid resolution | Emission scenarios |
|---|---|---|---|---|---|
| Australia | Commonwealth Scientific and Industrial Research Organization | CSIRO-MK3.0 | CSMK3 | 1.9° × 1.9° | SRA1B, SRB1 |
| Canada | Canadian Centre for Climate Modeling and Analysis | CGCM33.1 (T47) | CGMR | 2.8° × 2.8° | SRA1B |
| China | Institute of Atmospheric Physics | FGOALS-g1.0 | FGOALS | 2.8° × 2.8° | SRA1B, SRB1 |
| France | Centre National de Recherches Meteorologiques | CNRM-CM3 | CNCM3 | 1.9° × 1.9° | SRA1B, SRA2 |
| France | Institute Pierre Simon Laplace | IPSL-CM4 | IPCM4 | 2.5° × 3.75° | SRA1B, SRB1, SRA2 |
| Germany | Max-Planck Institute for Meteorology | ECHAM5-OM | MPEH5 | 1.9° × 1.9° | SRA1B, SRB1, SRA2 |
| Japan | National Institute for Environmental Studies | MRI-CGCM2.3.2 | MIHR | 2.8° × 2.8° | SRA1B, SRB1 |
| Norway | Bjerknes Centre for Climate Research | BCM2.0 | BCM2 | 1.9° × 1.9° | SRA1B, SRB1 |
| Russia | Institute for Numerical Mathematics | INM-CM3.0 | INCM3 | 4° × 5° | SRA1B, SRB1, SRA2 |
| UK | UK Meteorological Office | HadCM3 | HADCM3 | 2.5° × 3.75° | SRA1B, SRB1, SRA2 |
| | | HadGEM1 | HADGEM | 1.3° × 1.9° | SRA1B, SRA2 |
| USA | Geophysical Fluid Dynamics Lab | GFDL-CM2.1 | GFCM21 | 2.0° × 2.5° | SRA1B, SRB1, SRA2 |
| USA | Goddard Institute for Space Studies | GISS-AOM | GIAOM | 3° × 4° | SRA1B, SRB1 |
| USA | National Centre for Atmospheric Research | PCM | NCPCM | 2.8° × 2.8° | SRA1B, SRB1 |
| | | CCSM3 | NCCCS | 1.4° × 1.4° | SRA1B, SRB1, SRA2 |

**2.3. Scenarios**

Emission scenarios are estimates of future emissions of greenhouse gases and aerosols and are considered as the main inputs for climate models. Climate models use these estimates of future emissions to simulate future climate conditions. Determination of the exact amount of future emissions of these gases is not possible for future decades. Therefore, they are considered as one of the main sources of uncertainty in climate change projections and become more pronounced at long-term projections that simulate climate change conditions at the end of the 21$^{st}$ century (Stott and Kettleborough 2002; Hawkins and Sutton 2009).

The goal of this study was to investigate an alternative modeling approach for combining outputs of several climate models in order to reduce uncertainty in projections. In order to investigate the role of scenario uncertainty in projections, we considered three emission scenarios: A1B, A2, and B1. These scenarios were used in climate simulations by all selected 15 GCMs. Each scenario takes into account the dominant features of emissions of greenhouse gases such as physical, societal, and economic factors. In the present study we only made use of three scenarios briefly described as follows:

A1B: this scenario depicts a future world with balanced consumption across energy resources; a world with very rapid economic growth and rapid introduction of new and more efficient technologies, but with low population growth. Personal wealth is preferred over environmental quality in this world.

A2: this scenario depicts a differentiated world. In this scenario, high population growth, less concern for rapid economic development, strengthening regional cultural identities, with an emphasis on family values and local traditions are the underlying themes.

B1: this scenario depicts a convergent world. In this scenario, the introduction of clean technologies, rapid technology development, and movement towards achieving environmental and social sustainability are the underlying themes.

## 3. Methodology

### 3.1. Simulation of present climate

#### 3.1.1. Single model simulations

In some situations, a user usually has to decide beforehand which single model to choose for the decision-making process. This single model usually has a better performance than other models. In order to find the best single model for each variable in each station, we investigated which GCM would be more skillful in reproducing the variables in the present-day climate for the study region.

We used some known indices to evaluate the performance of the single models together with equally–weighted averaging of the model and the ANN combination approach in historical climate. In order to assess the skill of every single model in simulating monthly means of temperature and precipitation in historical climate for the study area, we used three indices for each scenario: coefficient of determination ($R^2$) (Eq. 1), index of agreement (IA) (Eq. 2), and root mean square errors (RMSE) (Eq. 3). $R^2$ and IA indicate the skill of the models in simulating the monthly means of the variables, and the more they are close to 1, the more it indicates that the monthly means of the simulations agree with observations. RMSE was used to investigate the accuracy of simulations of monthly means of the variables. RMSE is an error-index and demonstrates the bias between simulations and observations. This index has the same scale as the variables and therefore provides a good judgment for us about the range of bias in simulations. Each of the three indices does not represent the skill of the models in simulations individually, but taking all three indices together into consideration can tell us how skillful a model would be in simulating the historical climate.

$$R^2 = \frac{\left[\sum_{i=1}^{n}(S_i - \underline{S})(O_i - \underline{O})\right]^2}{\sum_{i=1}^{n}(S_i - \underline{S})^2 \sum_{i=1}^{n}(O_i - \underline{O})^2} \quad (1)$$

$$IA = 1 - \left[\frac{\sum_{i=1}^{n}(S_i - O_i)^2}{\sum_{i=1}^{n}(|S'_i| + |O'_i|)^2}\right] \quad (2)$$

$$RMSE = \left[\frac{1}{n}\left(\sum_{i}^{n}(Si - Oi)^2\right)\right]^{1/2} \quad RMSE = \left[\frac{1}{n}\left(\sum_{i}^{n}(S_i - O_i)^2\right)\right]^{1/2} \quad (3)$$

Where $S_i$ and $O_i$ are the *i*th simulation and observation, and $\underline{S}$ and $\underline{O}$ are the means of simulations ($S_i$) and ($O_i$), respectively. *n* is the total number of the evaluated samples. In Eq. (1-3) $S'_i$ and $O'_i$ are

$$S'_i = S_i - \underline{S} \quad (4)$$

$$O'_i = O_i - \underline{O} \quad (5)$$

In this section, the long-term monthly means of each variable were initially calculated from observations and GCM simulations. Then, the indices were calculated to compare the monthly means of the GCM simulations with observations in each station. The precipitation calculated by GCMs was based on mm/day which was changed to total precipitation in a month, based on mm, to match the observations.

### 3.1.2. Equally–weighted model averaging

Averaging the equally-weighted models or so-called the "mean model" is the simplest approach to combine outputs of several climate models and therefore to quantify uncertainty in projections (Lambert and Boer 2001; Tebaldi and Knutti 2007). Compared to single model simulations, this approach provides a better comparison to observations and is more straightforward than the weighting models based on their skill. We adopted the mean model approach as a reference to the best model and ANN approaches, and to see how much the mean model, in comparison with the two mentioned approaches, can reduce uncertainty in future climate simulations.

Outputs of the 15 employed GCMs were obtained from the Canadian Climate Data and Scenarios database (http://ccds-dscc.ec.gc.ca) for the baseline period of each variable and for each station. The baseline period for each station was defined based on the availability of observations in that station. First, the long-term monthly means of the variables were calculated from observations in each station. Then, the long-term monthly means of simulations were obtained from each GCM and each station based on its observation baseline. Finally, the long-term monthly means of simulations in each station were compared with their corresponding observations in each month. The comparison was made by using Eq. (1-3) and calculating the performance indices for the baseline period in each station.

### 3.1.3. The ANN combination approach

The objective of the present study was to investigate an alternative modeling approach to combining outputs of several climate model projections. We adopted the ANN approach to obtain a multi-model combination of multiple GCM projections, and to investigate how much this approach was able to improve projections. ANNs have been used in several climate studies (Karl et al. 1990; Trigo and Palutikof 1999; Sailor et al. 2000; Mpelasoka, Mullan, and Heerdegen 2001; R. Knutti et al. 2003; Boulanger, Martinez, and Segura 2006; Jean-Philippe Boulanger 2007). For instance, (Boulanger, Martinez, and Segura 2006; Jean-Philippe Boulanger 2007) used this approach to investigate future climate change conditions of temperature and precipitation over South America during the twenty-first century. They found that the ANN would underestimate the potential climate change projections simulated by the IPCC models.

The ANN has two main roles in this study. First, it obtains an optimal combination of several GCMs. The optimal combination in this method is calculated by the network itself based on the skill of climate models in simulating the present-day climate for the study region. Therefore, this method reduces the subjectivity and uncertainty aspects in constructing and combining metrics used for weighting the models. Second, the ANN approach correlates the GCM outputs on grid-scale to sub-grid scale possesses that are captured in observations on the local scale. GCMs lack any representation of the local environment especially the urban environment which may impact observations. The ANN approach provides a multi-model GCM projection which has been corrected for local environments especially for urban environments.

A detailed description of ANNs and multi-layer perceptron (MLP) can be found in numerous documentations in the literature. Therefore, in the present study, we will only focus on a brief summary of the methodology. The basic structure of every neural network involves inter-connected nodes that are arranged in layers. The architecture of every neural network is composed of an input layer, one or more hidden layers, and an output layer. Every node in the hidden and output layers consists of activation and transfer functions. Initially, in each node, the activation function value is calculated. Then the calculated value passes through a transfer function. This process is identical for all nodes in hidden and output layers. The input layer, however, does not contain any activation or transfer function and serves merely to transfer the inputs to the network. Finally, the output of the system is compared with the target value and the output error of the modeling system is calculated. The objective of the training phase is to reduce the output error of the modeling system to its minimum. In the back-propagation training algorithm, this task is accomplished by distributing the output error back into the system among network weights and adjusting the weights so that the final output error approximates the target value with a selected error goal.

Fig. 1 shows a schematic diagram of the used ANN architecture. In the present study, a three-layered feedforward MLP with a 15-30-1 network structure was used. The input layer consisted of 15 inputs which represented the monthly means of each GCM in the baseline period. Monthly means of the observations in each station were considered as the output of the network in the training phase. The historical simulated monthly means of each GCM were obtained from the Canadian Climate Data and Scenarios database (http://ccds-dscc.ec.gc.ca) for every station. A network with 15-30-1 node architecture was selected by trial and error and by considering the performance of each model architecture. Finally, 30 neurons were selected for the hidden layer because this number of nodes demonstrated the best performance in simulations. The dataset was divided into three subsets of the training set, test set, and validation set, each having 70%, 15%, and 15% of the total dataset, respectively. To evaluate the skill of each trained ANN, long-term monthly means of GCM simulations were given to the network and ANN values were obtained for each month. Then, the long-term monthly means simulated by the ANN were compared with their corresponding monthly observations by using Eq. (1-3). Finally, the skill of each trained ANN was evaluated by indices such as $R^2$, RMSE, and IA.

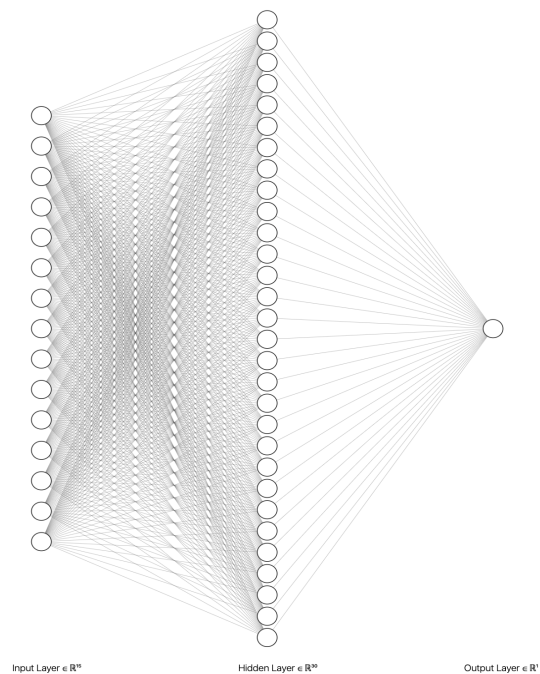

Input Layer ∈ ℝ$^{15}$    Hidden Layer ∈ ℝ$^{30}$    Output Layer ∈ ℝ$^{1}$

Fig. 1 schematic diagram of the used ANN architecture

## 3.2. 21st-century simulations

Simulation of the climate variables for the 21$^{st}$ century was conducted by using the trained ANNs that were developed for every station based on section 3.1.3. Projections of future monthly means of temperature and precipitation from each GCM for the stations in the study area were obtained from the Canadian Climate Data and Scenarios database (http://ccds-dscc.ec.gc.ca) database for the 2020-2100 period. The projections were based on A1B, A2, and B1 emission scenarios which were regarded as the input to the developed ANNs. The GCM monthly simulations for every station were given to the ANNs as their inputs and future monthly means of temperature and precipitation were projected by the developed ANNs in each station for the 2020-2100 period. Then, the simulated monthly means were averaged over every 20 years for four periods of 2020-2039, 2040-2059, 2060-2079, and 2080-2099 to demonstrate a better view of future changing trends in every station.

## 4. Results and discussion

### 4.1. Present-day climate simulations

As it was described in section 3.1.3, an ANN was developed for every variable and station. The statistics of the test phase of the developed ANNs are given in table 3. Moreover, the calculated performance indices of every single GCM are given in tables 4-7 for only the A1B emission scenario. A model that has the highest $R^2$ and D and the lowest RMSE can be considered as the best GCM for the study area. Among the 15 models, calculated $R^2$ and D are almost in the same range, and therefore RMSE could be regarded as the best index to distinguish the skilled GCM among the other models.

Table 3 Calculated performance metrics for the developed ANNs in the test phase

| Variable | Stations | No. of neurons | R | MAE | RMSE |
|---|---|---|---|---|---|
| Precipitation | Karaj | 30 | 0.56 | 14.92 | 24.02 |
| | Mehrabad | 30 | 0.56 | 16.93 | 27.66 |
| | Doshan Tappeh | 30 | 0.62 | 13.58 | 18.14 |
| | Abali | 30 | 0.68 | 22.33 | 28.33 |
| Temperature | Karaj | 30 | 0.96 | 2.204 | 2.885 |
| | Mehrabad | 30 | 0.98 | 1.382 | 1.721 |
| | Doshan Tappeh | 30 | 0.98 | 1.303 | 1.644 |
| | Abali | 30 | 0.98 | 1.773 | 2.128 |

Table 4 gives the calculated performance indices for temperature for every 15 GCM in this study. As table 4 indicates, there was not a unique single model that could be skillful over all 4 stations. Calculated indices indicate that at Doshan Tappeh, and Mehrabad stations, located in Tehran megacity, the MIROC3.2 medium resolution GCM had the best agreement between simulations and observations, and therefore the lowest uncertainty among the 15 models for Tehran megacity. Although this model did not have the highest $R^2$, it had the lowest RMSE among the single models which made it the best model for simulating temperature in the area. At Doshan Tappeh station, CSIROM and IPSL, and at Mehrabad station, CSIRO and GISS GCMs were the second and third models that had better skills in simulating the historical climate over the other models in the area, respectively. At Abali station, BCM2, and at Karaj station, the IPSLCM4 GCMs had the best agreement between simulations and observations, respectively. Therefore, these GCMs were considered as the best models for simulating temperature in these stations. At Abali station, NCAR, and CGCM3T, and at Karaj station, IPSL and ECHOG were the second and third models that had better skills in simulating the historical climate over the other models in the area, respectively.

Table 4 Validation of temperature simulated by single models

| Models | Abali- A1B | | | Doshan Tappeh-A1B | | | karaj- A1B | | | Mehrabad- A1B | | |
|---|---|---|---|---|---|---|---|---|---|---|---|---|
| | $R^2$ | IA | RMSE | $R^2$ | IA | RMSE | $R^2$ | IA | RMSE | $R^2$ | IA | RMSE |
| BCM2.0 | **_0.993_** | **_0.998_** | **_0.900_** | 0.980 | 0.746 | 9.206 | 0.982 | 0.876 | 6.237 | 0.975 | 0.747 | 9.151 |
| CGCM3T63 | 0.978 | 0.989 | 2.032 | 0.975 | 0.768 | 9.433 | 0.971 | 0.889 | 6.334 | 0.975 | 0.758 | 9.554 |
| CNRMCM3 | 0.978 | 0.972 | 2.988 | 0.990 | 0.853 | 7.138 | 0.983 | 0.949 | 4.050 | 0.989 | 0.845 | 7.230 |
| CSIROMk3.5 | 0.942 | 0.510 | 11.550 | 0.979 | 0.974 | 2.641 | 0.969 | 0.941 | 3.842 | 0.980 | 0.975 | 2.600 |
| ECHAM5OM | 0.977 | 0.748 | 8.674 | 0.997 | 0.933 | 4.420 | 0.993 | 0.990 | 1.636 | 0.998 | 0.935 | 4.349 |
| ECHO-G | 0.995 | 0.927 | 4.917 | 0.995 | 0.936 | 4.763 | 0.995 | 0.990 | 1.789 | 0.995 | 0.935 | 4.750 |
| GFDLCM2.1 | 0.979 | 0.901 | 5.409 | 0.988 | 0.935 | 4.486 | 0.981 | 0.989 | 1.792 | 0.987 | 0.937 | 4.384 |
| GISS-ER | 0.957 | 0.875 | 6.860 | 0.961 | 0.959 | 4.024 | 0.949 | 0.978 | 2.873 | 0.959 | 0.955 | 4.198 |
| HADCM3 | 0.943 | 0.932 | 4.566 | 0.975 | 0.899 | 5.734 | 0.963 | 0.971 | 2.974 | 0.979 | 0.900 | 5.670 |
| INMCM3.0 | 0.953 | 0.942 | 3.641 | 0.978 | 0.734 | 8.116 | 0.968 | 0.881 | 5.238 | 0.979 | 0.731 | 8.113 |
| IPSLCM4 | 0.995 | 0.875 | 5.871 | 0.984 | 0.938 | 4.249 | **_0.987_** | **_0.991_** | **_1.608_** | 0.984 | 0.937 | 4.244 |
| MIROC3.2 medres | 0.978 | 0.781 | 8.555 | **_0.987_** | **_0.993_** | **_1.551_** | 0.982 | 0.984 | 2.331 | **_0.987_** | **_0.993_** | **_1.578_** |
| MRI CGCM2.3.2a | 0.980 | 0.921 | 5.166 | 0.989 | 0.939 | 4.693 | 0.984 | 0.987 | 2.060 | 0.990 | 0.938 | 4.690 |
| NCARCCSM3 | 0.981 | 0.904 | 5.295 | 0.987 | 0.928 | 4.704 | 0.983 | 0.966 | 3.204 | 0.986 | 0.877 | 6.236 |
| NCARPCM | 0.972 | 0.991 | 1.590 | 0.991 | 0.680 | 10.050 | 0.984 | 0.833 | 7.021 | 0.991 | 0.675 | 10.078 |

Table 5 compares the calculated indices between the mean model and the ANN approach for temperature. As table 5 indicates, the mean model did not improve the simulations. The indices indicated that there were some single models that had better skills in simulating present-day climate than the mean model. However, there was a significant improvement in temperature simulations with the ANN approach. This approach considerably reduced RMSE and improved the temperature simulations by demonstrating the best skill compared to both the mean model and single models simulations.

Table 5 Validation of temperature simulated by the mean model and the ANN approach

| Method | $R^2$ | IA | RMSE | $R^2$ | IA | RMSE | $R^2$ | IA | RMSE | $R^2$ | IA | RMSE |
|---|---|---|---|---|---|---|---|---|---|---|---|---|
| SIMPLE AVE. | 0.984 | 0.915 | 5.127 | 0.992 | 0.919 | 5.104 | 0.987 | 0.985 | 2.142 | 0.992 | 0.918 | 5.122 |
| ANN | *0.992* | *0.998* | *0.892* | *0.999* | *1.000* | *0.345* | *0.981* | *0.995* | *1.262* | *0.998* | *0.999* | *0.459* |

Table 6 gives the calculated indices for every 15 GCM for precipitation. As table 6 indicates, similar to temperature simulations, there was not a single model that could be skillful over all four stations. At Mehrabad and Doshan Tappeh stations, located in the Tehran megacity, MRI and IPSL had the best agreement between simulations and observations, respectively. At Karaj and Abali stations, located near Tehran megacity, MRI and ECHO-G models had the best skill in simulating present-day climate, respectively.

Table 6 Validation of precipitation simulated by single models

| | Abali- A1B | | | Doshan Tappeh-A1B | | | Karaj- A1B | | | Mehrabad- A1B | | |
|---|---|---|---|---|---|---|---|---|---|---|---|---|
| Models | $R^2$ | IA | RMSE | $R^2$ | IA | RMSE | $R^2$ | IA | RMSE | $R^2$ | IA | RMSE |
| BCM2.0 | 0.30 | 0.65 | 37.73 | 0.21 | 0.00 | 50.51 | 0.36 | 0.00 | 50.51 | 0.25 | 0.05 | 55.21 |
| CGCM3T63 | 0.49 | 0.74 | 25.58 | 0.34 | 0.65 | 20.97 | 0.46 | 0.68 | 19.60 | 0.35 | 0.07 | 23.26 |
| CNRMCM3 | 0.04 | 0.51 | 33.26 | 0.01 | 0.02 | 34.25 | 0.04 | 0.00 | 35.28 | 0.00 | 0.00 | 38.02 |
| CSIROMk3.5 | 0.90 | 0.00 | 40.14 | 0.75 | 0.88 | 9.48 | 0.82 | 0.91 | 7.73 | 0.76 | 0.92 | 7.32 |
| ECHAM5OM | 0.75 | 0.56 | 37.84 | 0.46 | 0.70 | 12.54 | 0.60 | 0.77 | 10.76 | 0.49 | 0.79 | 11.85 |
| ECHO-G | *0.75* | *0.77* | *22.31* | 0.66 | 0.81 | 13.59 | 0.77 | 0.82 | 13.01 | 0.63 | 0.74 | 15.13 |
| GFDLCM2.1 | 0.70 | 0.61 | 28.98 | 0.59 | 0.87 | 11.12 | 0.75 | 0.92 | 8.67 | 0.66 | 0.09 | 9.16 |
| GISS-ER | 0.54 | 0.65 | 27.63 | 0.53 | 0.82 | 14.36 | 0.63 | 0.84 | 12.92 | 0.53 | 0.86 | 16.09 |
| HADCM3 | 0.75 | 0.29 | 34.52 | 0.61 | 0.80 | 11.52 | 0.80 | 0.82 | 7.51 | 0.70 | 0.88 | 8.43 |
| INMCM3.0 | 0.27 | 0.00 | 35.13 | 0.19 | 0.51 | 14.74 | 0.26 | 0.54 | 13.31 | 0.15 | 0.49 | 14.82 |
| IPSLCM4 | 0.85 | 0.57 | 29.09 | *0.91* | *0.97* | *4.99* | 0.83 | 0.95 | 6.27 | 0.93 | 0.97 | 5.46 |
| MIROC3.2 medres | 0.61 | 0.33 | 33.57 | 0.49 | 0.80 | 11.71 | 0.66 | 0.88 | 8.75 | 0.51 | 0.83 | 10.39 |
| MRI CGCM2.3.2a | 0.94 | 0.72 | 25.74 | 0.93 | 0.98 | 5.17 | *0.95* | *0.97* | *5.29* | *0.97* | *0.97* | *5.39* |
| NCARCCSM3 | 0.75 | 0.69 | 25.43 | 0.87 | 0.94 | 7.59 | 0.74 | 0.86 | 10.87 | 0.85 | 0.88 | 10.20 |
| NCARPCM | 0.69 | 0.00 | 41.03 | 0.59 | 0.53 | 15.77 | 0.74 | 0.65 | 13.15 | 0.63 | 0.67 | 12.56 |

Table 7 compares the calculated indices between the mean model and the ANN approach for precipitation. As table 7 indicates, the ANN approach did not have a satisfactory skill in simulating the historical period precipitation in all four stations. The ANN approach outperformed the single models in Abali and Mehrabad stations. However, In Doshan Tappeh and Karaj stations, single GCMs performed better than the ANN approach. For Doshan Tappeh and Karaj stations, the best model was an individual GCM (IPSL and MRI GCMs, respectively). Moreover, similar to temperature, the mean model did not improve the simulations in all 4 stations. There were some single models that had better skills in simulating historical precipitation than the mean model did.

Table7 Validation of precipitation simulated by the mean model and the ANN approach

| Method | $R^2$ | IA | RMSE | $R^2$ | IA | RMSE | $R^2$ | IA | RMSE | $R^2$ | IA | RMSE |
|---|---|---|---|---|---|---|---|---|---|---|---|---|
| SIMPLE AVE. | 0.62 | 0.59 | 27.39 | 0.49 | 0.78 | 13.53 | 0.66 | 0.82 | 11.94 | 0.51 | 0.72 | 14.79 |
| ANN | ***0.65*** | ***0.82*** | ***20.99*** | 0.87 | 0.86 | 9.47 | 0.82 | 0.95 | 6.36 | ***0.93*** | ***0.98*** | ***3.94*** |

Different calculated ranges of the indices such as RMSE and $R^2$ in the simulation of temperature and precipitation by single models indicate that the models can simulate temperature with higher confidence than precipitation in the present-day climate. Moreover, there is a substantial difference among single models in simulating the present-day precipitation. Unlike precipitation, the temperature has a narrower range of indices especially RMSE in simulations of present-day climate. These results are compatible with several studies such as (Hawkins and Sutton 2010) which have indicated this issue. A wider range of RMSE and $R^2$ in simulating baseline precipitation compared to temperature highlights the fact that models simulate the present-day precipitation with lower confidence than temperature. The low confidence in the simulation of precipitation is due to the fact that models are not able to correctly project some underlying sub-grid processes that influence precipitation change. Moreover, precipitation is strongly influenced by some local or regional geographic features such as mountainous terrain. These features are not usually well presented in current GCMs.

Furthermore, the identity of models and their ranking based on their skill changed between the two variables and among stations, i.e. there was not a unique model which could represent the best model for all variables and/or stations over the region. These results are similar to results from studies such as (Hagedorn, Doblas-Reyes, and Palmer 2005; Gleckler, Taylor, and Doutriaux 2008) which indicated that the models were best for temperature were not necessarily best for other variables such as precipitation. The mean model which was calculated from simple averaging the outputs of the single models was also considered in this study as a reference method for the ANN combination approach. As the calculated indices indicated, the mean model only provided the mean state of a variable and did not agree well with the present-day climate compared to some single model simulations and the ANN combination approach.

Compared to the mean model, the indices indicated that the ANN combination approach significantly improved the simulations of present-day climate. The ANN combination approach improved the IA and $R^2$ and considerably reduced the RMSE, especially in temperature simulations. In simulating temperature, the ANN approach demonstrated to have the best skill at simulating present-day monthly means of the variables than the mean model and the best model in all 4 stations. In simulating the present-day precipitation, however, the ANN approach was not the best approach in all stations although it performed better than the mean model. In Abali and Mehrbad stations the ANN had the best skill in simulating the historical precipitation. In Doshan Tappeh and Karaj stations, however, single GCMs had better skills than the other two approaches and were the best single models for simulating the precipitation. The reason for the better performance of some single models over the ANN combination approach in simulating the present-day precipitation in some stations maybe because some single models may resolve the sub-grid processes in simulating the precipitation such as the geographical features of the study location better than other GCMs do. Moreover, we used all available models to incorporate all skills of the models into multi-model simulations. In a multi-model approach based on present-day skills of models, due to the low skill of some models in simulating a variable, some models affect the outcome of a multi-model projection by reducing the accuracy of simulations (Giorgi and Mearns 2002; Tebaldi and Knutti 2007). In addition, a study by (Hagedorn, Doblas-Reyes, and Palmer 2005) showed that for some variables, the multi-model combination might not be significantly better than the best single model. He concluded that the performance of a multi-model combination approach must be evaluated when considering its overall performance over all aspects of predictions.

To sum up, the results indicated that the proposed ANN combination approach to combining GCM simulations is able to reduce uncertainties and therefore to improve reliability in climate projections, especially for temperature compared to the best single model and the simple averaging approach. Therefore, based on its performance in present-day climate, the ANN approach was adopted to produce a multi-model projection of temperature and precipitation for the study area.

### 4.2. 21$^{st}$ century projections

The performance of the ANN combination approach based on simulating temperature and precipitation for the present-day climate was investigated in section 4.1. The ANN combination approach demonstrated to have a better skill over the mean model and the best single model in combining outputs of present-day climate models and delivering more reliable results in all stations. Therefore, in order to investigate the future changes in temperature and precipitation in the study region, we used the ANN approach to provide a multi-model projection of the variables by combining projections from 15 GCMs for the 21$^{st}$ century for the study region.

Fig. 2 illustrates the projected temperature change for every station. Mehrabad and Doshan Tapeh stations are located in Tehran megacity and usually have higher temperatures. Karaj station is located in the Karaj urban area on the left side of Tehran megacity, and Abali station is located on the heights of Abali with usually lower temperature and higher precipitation than the other three stations. Projections of future climate conditions by the ANN multi-model approach indicated an increase of temperature in all stations and for all scenarios, even in Abali station where usually has lower temperatures due to its higher altitude. Comparing the three scenarios (A2, A1B, and B1) showed that the projected patterns were similar in all stations and differed mainly in their amplitude. Among the stations, projections suggested that Abali station would experience the least warming of about 1-2 $^{0}$C, and Doshan Tappeh station would experience the largest warming of about 3-4 $^{0}$C among all scenarios at the end of the 21$^{st}$ century. Moreover, the projected changes in temperature were greater for stations located in Tehran megacity than stations in its neighboring areas, like Karaj and Abali stations. This may be because the ANN approach is capable of incorporating the effect of the urban environment into the projections. Therefore, the coarse resolution GCM outputs for the study region are corrected for the Tehran urban environment by establishing a relationship between baseline simulations and observations. Furthermore, as (Stott and Kettleborough 2002) noted, the contribution of scenario uncertainty to projections would increase for lead times more than 30 years. As the multi-model projections indicate, differences among scenarios became more pronounced in the second part of the 21$^{st}$ century which is compatible with similar results such as those (Stott and Kettleborough 2002) and (Hawkins and Sutton 2009). The scenarios departed from each other in projections after the first period (2020-2039) and the divergence grew among the scenarios up to the end of the 21$^{st}$ century. A2 was the scenario with the greatest increase and B1 was the scenario with the smallest increase at the end of the century in all stations.

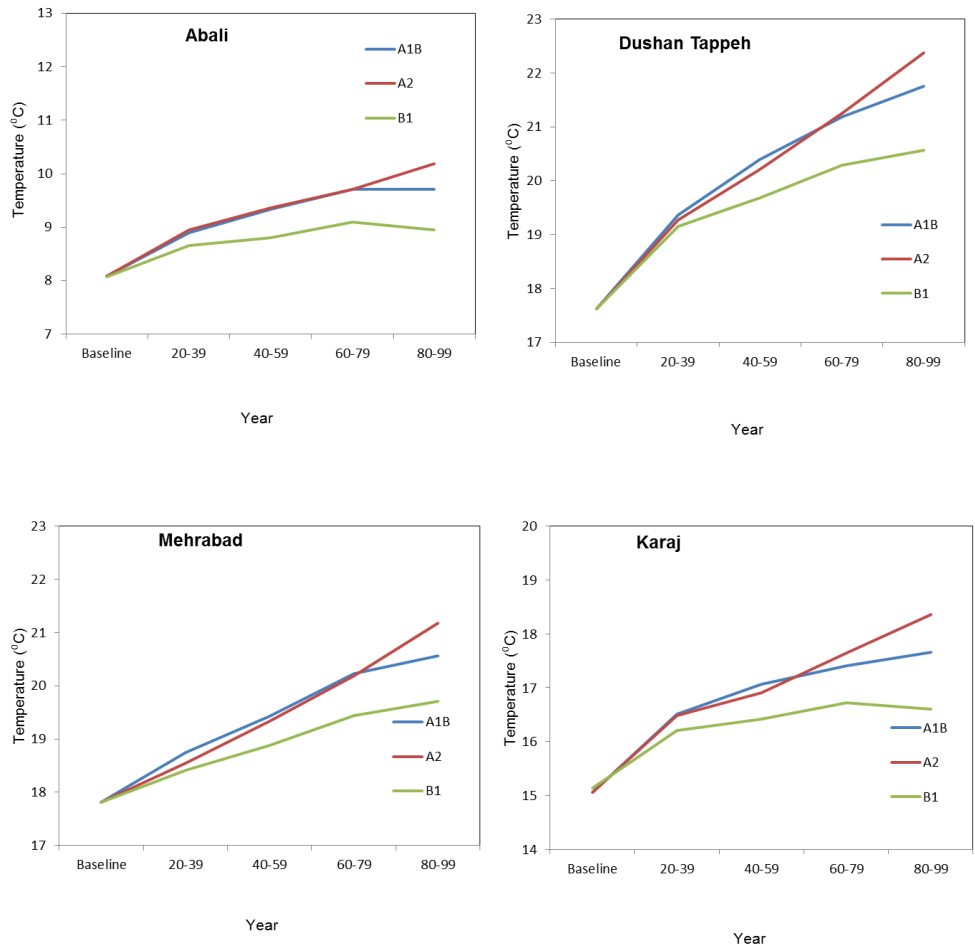

Fig. 2 Multi-model projection of temperature by the ANN combination approach

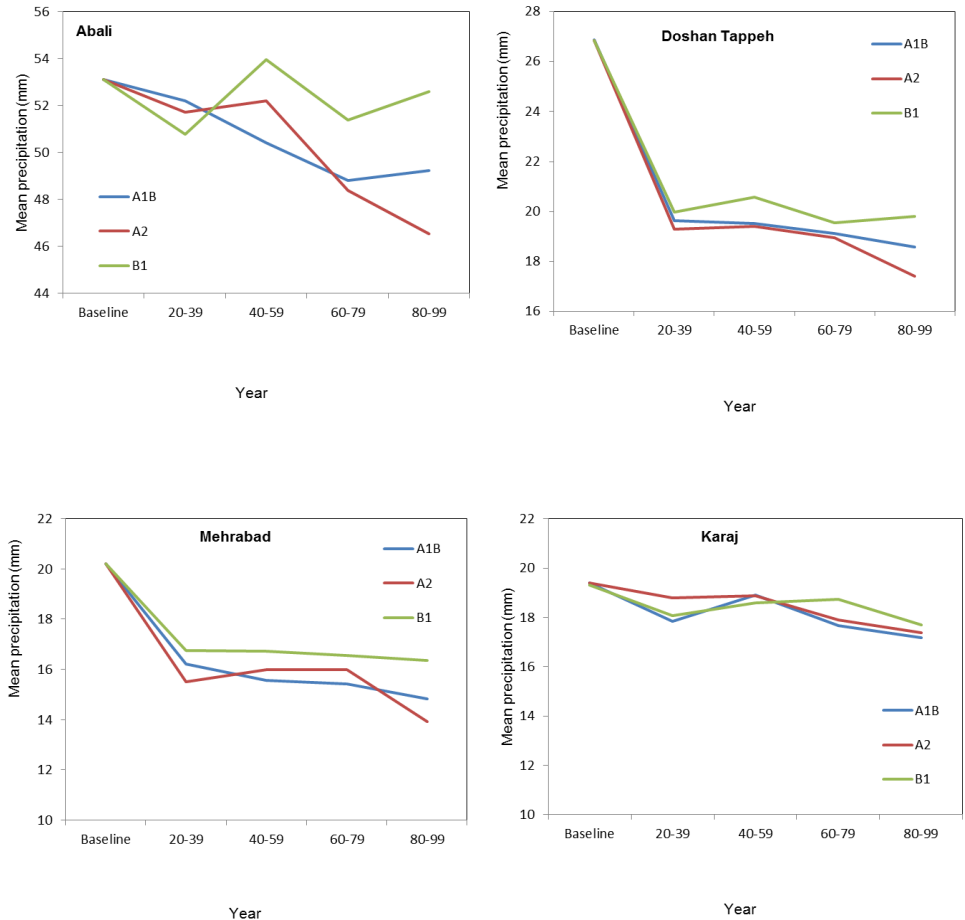

Fig. 3 Multi-model projection of precipitation by the ANN combination approach

Fig. 3 illustrates the future changes in precipitation for every station. The ANN approach projected a decrease in precipitation in all stations and for all scenarios. Comparing the three scenarios (A2, A1B, and B1) showed that the projected patterns were similar in all stations and differed mainly in their amplitude. Among the stations, ANN projections indicated that the Karaj station would experience the least reduction of about 1.5-2 mm among all scenarios at the end of the 21$^{st}$ century. Similar to temperature, the Doshan Tappeh station experienced the largest changes. The ANN projected the greatest reduction of about 7-9.5 mm at the end of the 21$^{st}$ century. Climate models represented general patterns of temperature fairly better than precipitation. Among stations, projections were more uncertain in Abali station. Projections had greater amplitude in Abali than in other stations. In the long-term, B1 did not indicate any reductions in this station and diverged from the other two scenarios since the first period. However, A1B and A2 projected a decrease in precipitation similar to other stations. Similar to temperature projection, scenarios departed from each other in projections after the first period and the uncertainty grew among the scenarios in the second part of the 21$^{st}$ century. Moreover, A2 projected the largest decrease and B1 projected the smallest decrease in precipitation at the end of the 21$^{st}$ century in all stations.

## 4. Conclusion

The goal of this study was to investigate whether combining model projections by ANN combination approach could improve multi-model projections and therefore reduce the uncertainty in climate projections. In order to establish a reference for the ANN combination approach, the equal weighting of the models (the mean model) and single climate models (the best single model) was also considered in the study.

The present study showed that the ANN combination approach was successful in combining outputs of several climate models and in reducing the uncertainty in simulations of present-day climate variables. Based on the calculated performance indices for the three approaches, it was concluded that projections based on single model simulation might not yield reliable results because single model simulations showed that the identity of models and their ranking based on their skill changed between the two variables and also among stations. The mean model was also not skillful in giving a reliable projection of present-day climate. However, calculated performance indices indicated that combining model projections by the ANN approach significantly improved the simulations of present-day temperature and precipitation than the single model and the mean model approaches. Based on the present-day skill of each approach, it was concluded that the ANN approach could give the best estimate of future trends of temperature and precipitation for a local environment. Therefore, the ANN approach was used to obtain projections of future temperature and precipitation for the study area.

The ANN approach used in this study can benefit the climate change projections due to the fact that it derives an optimal combination of several climate models by correlating the GCM simulations at grid-scale to observations of climate variables at the local scale. Therefore, this procedure reduces the subjectivity and uncertainty aspects in constructing and combining metrics used for weighting the models and delivers a multi-model projection that has been corrected for a specific local environment, especially for urban environments. However, the ANN approach is subject to some limitations which exist in similar skill-based performance studies of models. The optimal combination of models is derived based on the skill of the models in simulation of the historical climate. The underlying assumption governing this approach is the stationary relation between observed and simulated trends. This relation is formed in the training period of the ANN based on the twentieth-century climate and is applied to future simulations. A debate that exists here is that the skills of climate models are evaluated based on their performance in present-day climate conditions, and it is likely that the present optimal combination of models may not be the optimal combination in the future climate. This issue is due to some limitations that exist among present models. For instance, some characteristics of the climate models such as model parameterizations, or impacts of some physical processes such as carbon cycle feedbacks may change under future climate forcing (Frame et al. 2007; Reto Knutti et al. 2009). However, the only guidance that we have to evaluate the performance of current models is to evaluate their skills by comparing their simulations against observations of different present-day climate aspects. We might not be able to judge whether the closest projection to a multi-model average of future projections would be the best estimate of future climate due to the mentioned limitations, but for the present-day climate, we can decide that if a methodology gives better simulations of different aspects of present-day climate compared to observations, it would be a more skillful methodology and might give more reliable results for present climate. Consequently, using the skills of models based on their present-day performance may be a good measure for constructing a multi-model combination of models. The difficulty remains in how to integrate the present-day skills of models into their future projections. The effort of this study was to address this issue by correlating multiple climate models' projections to climate observations at a local station, but the methodology is subject to some limitations. Therefore, as many studies such as (Reto Knutti et al. 2009) have concluded, future research would benefit from developing methodologies to select and weight models and developing new approaches to combine multi-model projections to assess and reduce uncertainty in future climate projections.